\documentclass[12pt,a4paper,dvips]{article}
\usepackage{pic03}
\usepackage{url}
\usepackage{graphicx}

\begin{document}

\title{
\textbf{NEW CKM-RELATED STUDIES ON \textit{b} DECAYS 
IN THE DELPHI EXPERIMENT AT LEP}
}
\author{
Winfried A. Mitaroff \\
\textit{Institute of High Energy Physics,
Austrian Academy of Sciences, Vienna} \\
(on behalf of the DELPHI Collaboration)
}
\maketitle

\vspace{-1.5cm}
\baselineskip=14.5pt
\begin{abstract}
The $e^{-}e^{+}$ collider LEP, running at $\sqrt{s} = m(Z^{0})$, 
has been a copious source of $b$\,-hadrons produced in decays 
$Z^{0} \rightarrow b \, \bar{b}$. We present recent studies 
using up to $4 \times 10^6$ hadronic $Z^{0}$ decays 
acquired by the DELPHI detector between 1992 and 2000. 
They rely on efficient particle identification, precise 
track and vertex reconstruction and sophisticated data analysis 
algorithms.

Presented are: a new measurement of the CKM matrix element 
$| V_{cb} |$ in the semileptonic exclusive decays 
$\bar{B}^{0}_{d} \rightarrow D^{*+} \ell^{-} \bar{\nu}_{\ell}$; 
a new measurement of the $B^{0}_{d} - \bar{B}^{0}_{d}$ 
oscillation frequency $\Delta m_{d}$; 
and searches by three methods for 
$B^{0}_{s} - \bar{B}^{0}_{s}$ oscillations, yielding new 
lower limits on $\Delta m_{s}$.

\end{abstract}
\baselineskip=17pt

\section{
$| V_{cb} |$ from s.l. exclusive decays 
$\bar{B}^{0}_{d} \rightarrow D^{*+} \ell^{-} \bar{\nu}_{\ell}$ 
}

This analysis is performed on the exclusive channels 
$\ell^{-} = e^{-}$ or $\mu^{-}$, 
$D^{*+} \rightarrow D^{0} \pi^{+}$, 
% $D^{0} \rightarrow K^{-} \pi^{+}$ (1) or 
% $K^{-} \pi^{+} \pi^{+} \pi^{-}$ (2) or 
% $K^{-} \pi^{+} (\pi^{0})$ (3) \footnote{
$D^{0} \rightarrow K^{-} \pi^{+}$ or 
$K^{-} \pi^{+} \pi^{+} \pi^{-}$ or 
$K^{-} \pi^{+} (\pi^{0})$ \footnote{
\,the charge-conjugate states 
($B^{0}_{d} \rightarrow D^{*-} \ell^{+} \nu_{\ell}$, 
$D^{*-} \rightarrow \bar{D}^{0} \pi^{-}$, 
$\bar{D}^{0} \rightarrow K^{+} \pi^{-} \ldots$) 
are implicitly considered as well. }
by measuring the differential partial width (i.e. decay rate) 
which is, according to HQET, given by
\begin{equation}
\frac{\mathrm{d} \Gamma}{\mathrm{d} \omega} = 
\frac{G_{F}^{2}}{48 \pi^{3}} 
\cdot \mathcal{K} ( \omega ) 
\cdot \mathcal{F}_{D^{*}}^{2} ( \omega ) 
\cdot | V_{cb} |^{2}
\label{equ1}
\end{equation}
as a function of the $D^{*}$ boost $\omega$ in the $B^{0}_{d}$ 
rest frame, defined as
\begin{equation}
\omega ( q^{2} ) \equiv v_{B^{0}} \bullet v_{D^{*}} = 
\frac{m_{B^{0}}^{2} + m_{D^{*}}^{2} - q^{2}}
{2 \, m_{B^{0}} \, m_{D^{*}}}, \qquad
q^{2} \equiv ( p_{B^{0}} - p_{D^{*}} )^{2}
\label{equ2}
\end{equation}

\begin{figure}[htbp]
\centerline{ \hbox{
\hspace{0.2cm}
\includegraphics[width=70mm]{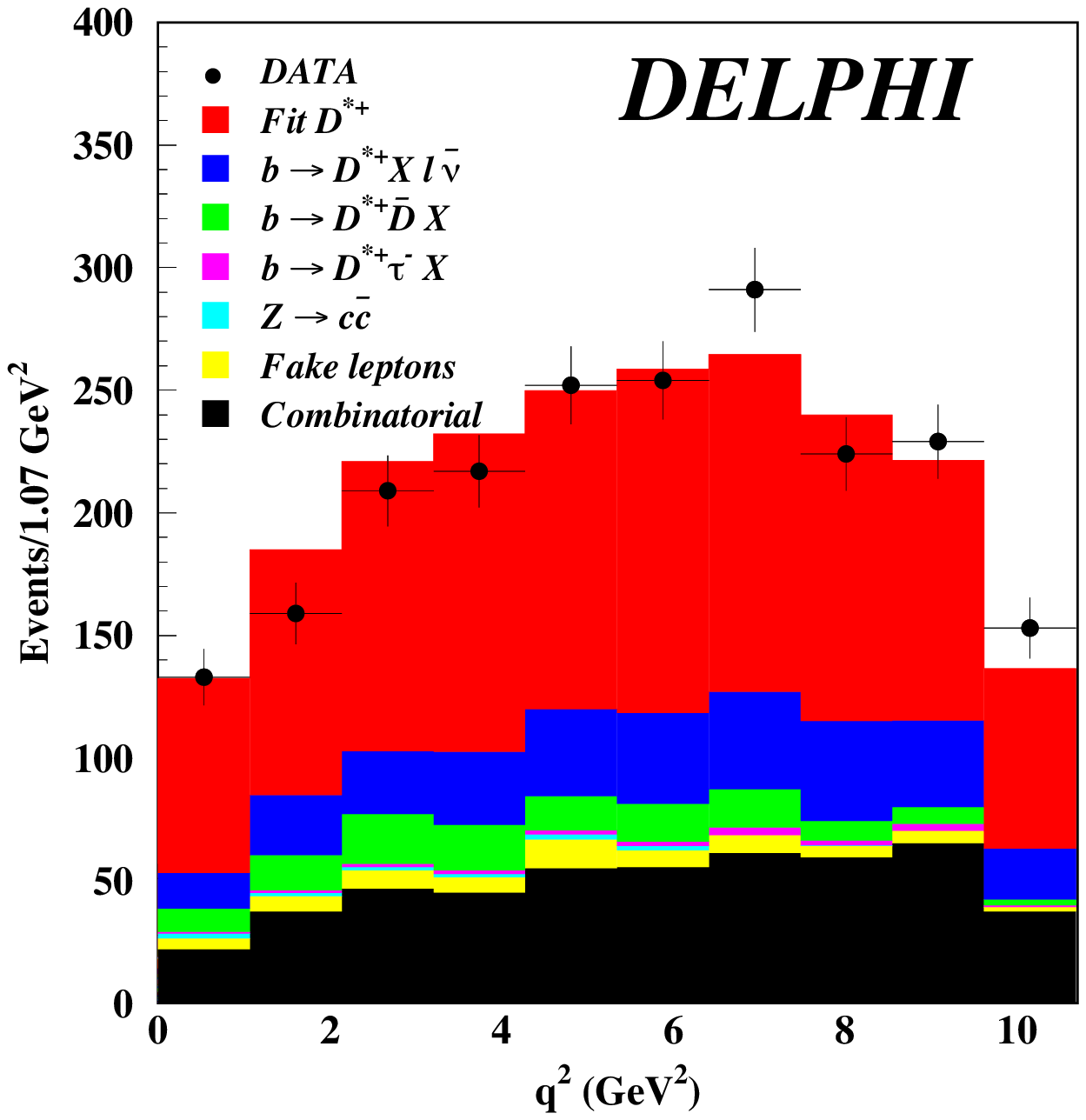}
\hspace{0.3cm}
\includegraphics[width=70mm]{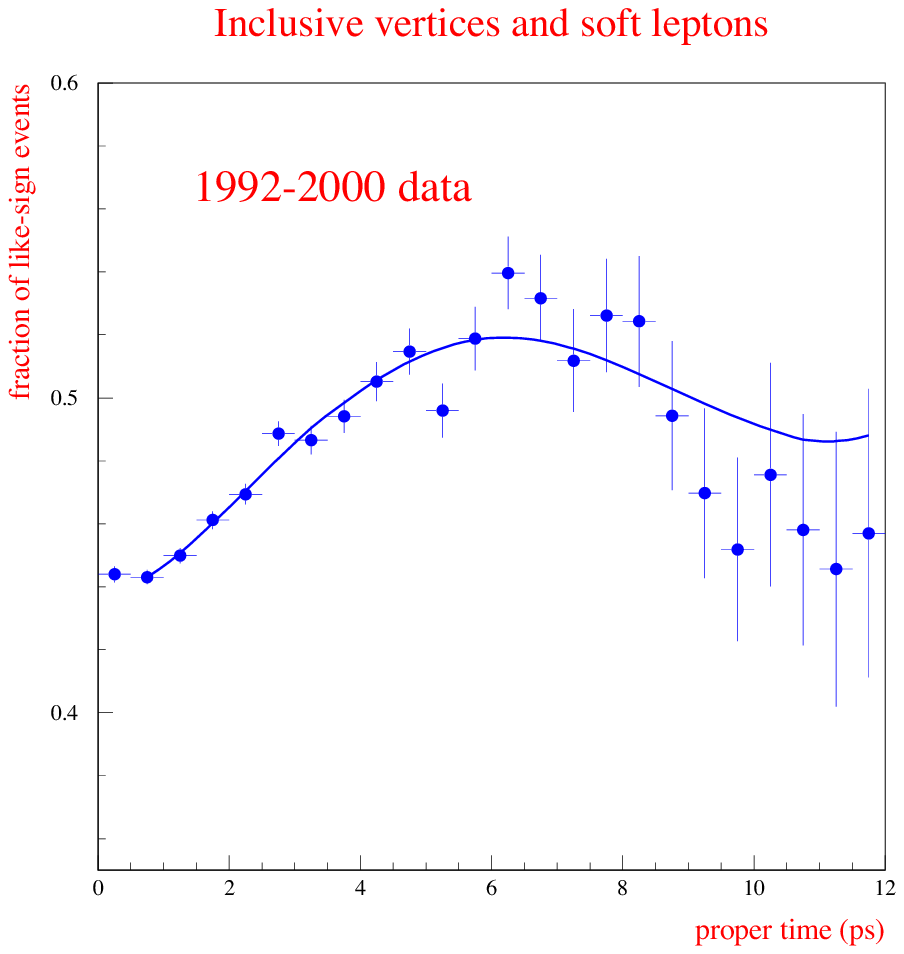}
} }
\caption{
(a) \textit{$| V_{cb} |$ analysis: 
distribution of $q^{2}$ for $D^{*}$ candidate events  (dots) 
with their fitted contributions (shaded).}
(b) \textit{$\Delta m_{d}$ analysis: 
fraction of like-sign tagged events as a function of the 
reconstructed proper time (data and fit).}
\label{twofigs}
}
\end{figure}

Its range is $1 \le \omega \; { _{\sim}^{<}} \; 1.5$, 
with the lower bound corresponding to $D^{*}$ zero recoil. 
$\mathcal{K} ( \omega )$ is a known kinematic factor, and
$\mathcal{F}_{D^{*}} ( \omega )$ is the hadronic form factor 
which may be expanded at $\omega = 1$. 
% \begin{equation}
% \mathcal{F}_{D^{*}} ( \omega ) = \mathcal{F}_{D^{*}} (1) 
% \cdot \left[ 1 - \rho_{\mathcal{F}}^{2} ( \omega - 1 ) 
% + \mathcal{O} ( \omega - 1 )^{2} \right]
% \end{equation}
$\mathcal{F}_{D^{*}} (1) \! \cdot \! | V_{cb} |$ and 
$\left[ \mathrm{d} \mathcal{F_{D^{*}}} / 
\mathrm{d} \omega \right]_{\omega = 1}$ 
are fitted from data, using eq. (\ref{equ1}) convoluted 
with the experimental resolution as a function of $q^{2}$, 
and extrapolated to $\omega \rightarrow 1$. 
Since $\mathcal{K} (1) = 0$, 
a reasonably constant reconstruction efficiency is required 
at $\omega \approx 1$.
% The data sample is based on 
% $3.37 \times 10^6$ hadronic $Z^{0}$ decays of 1992 -- 95 
% (reprocessed event reconstruction).
% Numbers of selected candidates are 
% $521 \pm 31$, $387 \pm 31$ and $780 \pm 51$ 
% for channels (1), (2) and (3), respectively. 
Separation of different decay mechanisms producing 
$D^{*}$ in the final state (notably for the 
exclusion of $D^{**} \rightarrow D^{*} X$ background) 
is achieved by novel algorithms 
and is shown in fig. {\ref{twofigs}}a.

Results of the fit, a calculation of $| V_{cb} |$ using 
$\mathcal{F}_{D^{*}}(1) = 0.91 \pm 0.04$ \footnote{
\,in the heavy quark limit, $\mathcal{F}_{D^{*}}(1) \rightarrow 1$; 
non-pertubative QCD corrections yield the value cited. }, and the 
corresponding decay branching fraction are {\cite{Vcb_slexcl_03}}: 
\begin{eqnarray*}
\mathcal{F}_{D^{*}} (1) \cdot | V_{cb} | & = & 
( 38.0 \pm 1.8 \pm 2.1 ) \times 10^{-3}                 \\
| V_{cb} | & = & 
( 41.8 \pm 2.0 \pm 2.3 \pm 1.7_{theor} ) \times 10^{-3} \\
% \rho_{\mathcal{F}}^{2} & = & 1.32 \pm 0.15 \pm 0.33     \\ 
\mathcal{BR} (\bar{B}^{0}_{d} \rightarrow 
D^{*+} \ell^{-} \bar{\nu}_{\ell}) & = & 
(5.54 \pm 0.20 \pm 0.41) \% 
\end{eqnarray*}

\vspace{-0.5cm}

\section{
Studies of $B^{0}_{d} - \bar{B}^{0}_{d}$ and 
$B^{0}_{s} - \bar{B}^{0}_{s}$ oscillations
}

Mixing of $B^{0}_{q} - \bar{B}^{0}_{q}$ ($q = d$ or $s$) 
proceeds via $2^{nd}$ order weak transitions (box graphs) 
which are dominanted by $t$-quark exchange.  
The probabilities $\mathcal{P}^{\, mix}_{nomix}$ of a 
$B^{0}_{q}$ ($\bar{B}^{0}_{q}$) to have, 
after some time $t$, mixed or not mixed into a 
$\bar{B}^{0}_{q}$ ($B^{0}_{q}$) state are 
\begin{equation}
\mathcal{P}^{\, mix}_{nomix} = \frac{1}{2 \, \tau_{q}} 
\cdot e^{- \frac{t}{\tau_{q}}} \cdot 
\left[ \mathrm{cosh} \, \frac{\Delta \Gamma_{q}}{2} \, t 
\; \mp \; \mathrm{cos} \, \Delta m_{q} \, t \right]
\label{equ3}
\end{equation}
The SM predicts $\Delta \Gamma_{q} \, \ll \, \Delta m_{q}$, 
thus the $\mathrm{cosh}$ term is approximated by 1. 

The oscillation frequencies $\Delta m_{d}$ and $\Delta m_{s}$ 
are directly related to $| V_{td} |$ and $| V_{ts} |$, 
respectively. 
Their measurements in a time-dependent analysis rely on 
two basic requirements: precise measurement of the proper 
decay time of the $B$ meson, achieved by precise track momentum 
and vertex reconstruction; and efficient tagging of the $B$ 
meson's flavour, both at production and decay.
% Time-dependent oscillations of $B^{0}_{d} - \bar{B}^{0}_{d}$ 
% have first been measured at LEP and later by BaBar and BELLE. 
% None have been observed so far in $B^{0}_{s} - \bar{B}^{0}_{s}$. 
% Here we present two recent updates of earlier studies.

\subsection{
$\Delta m_{d}$ from $B^{0}_{d} - \bar{B}^{0}_{d}$ oscillations
}

% The data used in this analysis are about $4 \times 10^6$ hadronic 
% $Z^{0}$ decays of 1992 -- 2000 (reprocessed event reconstruction).
% Selected candidates are $770 \times 10^3$ events, 
% $155 \times 10^3$ of which contain a soft lepton.
A high-statistics analysis, based on inclusive secondary vertex 
reconstruction and fitting $\Delta m_{d}$ (fig. {\ref{twofigs}}b) 
as well as an upper limit of $| \Delta \Gamma_{d} | / \Gamma_{d} $ 
{\cite{Osc_ds_ivx_02}}: 
\begin{eqnarray*}
\Delta m_{d} & = & 0.531 \pm 0.025 \pm 0.007 \, \mathrm{ps}^{-1} \\
| \Delta \Gamma_{d} | / \Gamma_{d} & < & 0.18 
\; \mathrm{at \, 95\% \, c.l.} 
\end{eqnarray*}

\subsection{
Search for $B^{0}_{s} - \bar{B}^{0}_{s}$ oscillations
}

(a) An analysis using the same method as that for $\Delta m_{d}$ 
(see section 2.1 above) {\cite{Osc_ds_ivx_02}}; 
and two analyses using new sophisticated algorithms, based on 
(b) inclusive high-$p_{t}$ leptons or (c) reconstructed 
$\bar{B}^{0}_{s} \rightarrow D_{s}^{+} \ell^{-} \bar{\nu}_{\ell} X$ 
events {\cite{Osc_s_hipt_03}} yield: 
\begin{eqnarray*}
\mathrm{(a)} \qquad \Delta m_{s} & > & 5.0 \, \mathrm{ps}^{-1} 
\; (\mathrm{sensitivity = 6.6 \, ps}^{-1}) 
\; \mathrm{at \, 95\% \, c.l.}                                 \\
\mathrm{(b)} \qquad \Delta m_{s} & > & 8.0 \, \mathrm{ps}^{-1} 
\; (\mathrm{sensitivity = 9.1 \, ps}^{-1}) 
\; \mathrm{at \, 95\% \, c.l.}                                 \\
\mathrm{(c)} \qquad \Delta m_{s} & > & 4.9 \, \mathrm{ps}^{-1} 
\; (\mathrm{sensitivity = 8.6 \, ps}^{-1}) 
\; \mathrm{at \, 95\% \, c.l.}
\end{eqnarray*}

% \section{
% Comparison with world averages {\cite{HFAG_summer03}}
% }
%
% \begin{center}
% $\mathcal{F}_{D^{*}} (1) \cdot | V_{cb} | $ = $38.8 \pm 1.1$ 
% ($\chi^{2}/ndf = 20.5/14$)
%
% $\Delta m_{d}$ ($B^{0}_{d} - \bar{B}^{0}_{d}$) = 
% $0.502 \pm 0.006$ ps$^{-1}$
%
% $B^{0}_{s} - \bar{B}^{0}_{s}$ osc. amplitude 
% (at $\Delta m_{s}$ = 15 ps$^{-1}$) = $0.51 \pm 0.40$
% end{center}
%
\section*{Acknowledgements}
Thanks are due to Franz Mandl (HEPHY Vienna) for his assistance 
in the poster presentation session at the conference.

\baselineskip=17pt

\end{document}